# Using Digital Twins and Intelligent Cognitive Agencies to Build Platforms for Automated CxO Future of Work

John-Thones Amenyo

Department of Mathematics and Computer Science

York College, CUNY

jtamenyo@york.cuny.edu

**Abstract**

AI, Algorithms and Machine based automation of executive functions in enterprises and institutions is an important niche in the current considerations about the impact of digitalization on the future of work. Building platforms for CxO automation is challenging. In this paper, design principles based on computational thinking are used to engineer the architecture and infrastructure for such CxO automation platforms.

*Keywords:* future of work, AI, robots, drones, human cognitive augmentation, digital twin, intelligent cognitive agencies, intelligent cognitive assistants, digitalization, disruptive impact of automation, CxO automation.

**Introduction**

There is great interest to research and discuss challenges, concerns and aspects of Future of Work, due to the disruptive impacts on human labor, employment, careers, and professions, of advances in digital technologies such as automation, AI (artificial intelligence), machine learning, robotics, drones, etc. Most of the current focus is on the replacement of human labor in various jobs and occupations, by substituting intelligent, smart, knowledge-based machines. Several topics and themes addressed in such discussions include: job automation; human labor, asset and capital replacement in jobs; human augmentation, assistance, prosthesis with intelligent cognitive assistants, bots, agents; human-computer (or machine, AI, robots, drones) collaboration and cooperation.

It is widely accepted and anticipated that advanced digital technology will have tremendous impact and consequences for the future of work within the next decade (10 to 20 years), (Balliester & ElSheikhi 2018, World Economic Forum, 2016).

The advances in technology which are the driving forces include various aspects of digital automation, digital intelligence, robots and robotics, drones, digital autonomous systems, autonomous systems, autonomic (self-*) systems, AI, Algorithmic Intelligence, Machine Intelligence, data-driven intelligence and automation.

All concerns, aspects and roles of work, jobs, employment can be expected to be affected, disrupted and changed, including, labor and workforce, work place, work practices, work culture, work context, work location, work compensation, work communities, work organization, work processes, work products, work goals, work motivators, work facilities, equipment and instruments, work specialization and divisions of labor, work roles and work habits. What is not known at this time is the exact nature and details of the impact, how deep, extent, degree and how widespread the effects, including job losses, new job creation, and work force augmentation and adaptation.

This paper is a first report regarding the Architecture Engineering aspects of a Project, (the CxO Automation FOW Platform Project), that focusses on one part of Future of Work that has not been sufficiently addressed so far. The objectives and key results (OKR) of the Project is concerned with ***how to build, deploy, operate and manage digital platforms that can automate the functions, tasks and jobs of C-suites or CxO's, for machines and algorithms to run large-scale enterprises (LSE), multi-national corporations (MNC), and global worldwide companies (GWC)***. The CxO (also known as C-suite) roles being targeted include business leaders such as, CEO, CFO, CHRO, CTO, CIO, CDO, CMO, etc., [as well as roles belonging to executive management, senior management, upper management, and board of directors], with specific focus on these roles in large-scale enterprises (LSE) and multinational corporations (MNC) with global scope and distributed operations. The rationale for this choice of focus is that the strategic, tactical and operational decisions at the CxO levels and roles have widespread immediate and long term impact a lot of corporate stakeholders, including employees, investors, communities, municipalities and governments. So, it is of great interest to research and investigate how advanced digital technologies will impact, change and disrupt the future of work in the CxO roles. Furthermore, one manifestation of digital and automation technologies and the internet and online driven rapid creation of global scale companies (Amazon, Google, EBay, Uber, Airbnb, Facebook, Alibaba, Baidu). Even now there is limited opportunity for aspirants and students, (especially, for women, minorities and members of disadvantaged groups), to learn and train to become CxO's for MNCs and LSEs. This situation will only be aggravated in the era of digital technology impact on future of work. Thus, there is an urgent need for what the CxO Automation FOW Platform Project aims to tackle: to research, develop and provide tools for studying and analyzing technology impacts on future work for CxO's.

Traditional FOW concerns are about O(100,000) thousands of job types disappearing, and throwing O(10 million) to O(100 millions) of humans out of work or into underemployment. Comparatively few CxO jobs are involved and available worldwide, O(10,000); however, the economic, political and cultural decisions made at the CxO level affect literally O(1 million) to O(100 million) workers, employees, dependents, families and communities.

The paper discusses the engineering of a technology-powered platform of organized complex of intelligent agencies for automating CxO functions. The focus is how to specify, develop and deploy a modular, reconfigurable and field programmable, intelligent digital platform that can be used by existing and potential employees of corporations (businesses, companies, or enterprises).

Besides addressing the technical challenges of using machine & intelligent algorithms teams and collectives to automate, run and manage enterprises and institutions, variants of such CxO Automation platforms have other future uses:

- Train future CxO's – the equivalent of (aircraft) flight simulators and learning environments and sandboxes.
- Augment CxO's knowledge, skills, capabilities, and expertise, by providing digital and cognitive prostheses and exoskeletons.
- Provide insight about CxO roles and functions to non CxO's and outsiders. Cognition Augmentation focus: Learning; (Non-formal, practical) & Lifelong Learning. Intended Learners,

Users (Who is to Learn): Higher Education Students, Employees of MNC-GWC-LSE; Employees of SME (Tech) Start Ups. (MNC: multinational corporations; GWC: global worldwide corporations; LSE: large-scale enterprises; SME: small and medium enterprises).

- Use as basis or foundation for Serious Games; Simulations; and Synthetic Data Generation.

*Providing insight about CxO functions to outsiders.*

The learners and students can use the platform to explore, discover, understand, and gain insight, familiarity, experience and expertise about human cognitive augmentation that enable human employees to survive in future corporate work environments based human-machine partnerships that result from the impact of digital and advanced automation, intelligent technology on the future roles and practices, in the future of work.

Learning Content (What to learn about): New, advancing technologies; new, incoming work environments and workspaces (which are digital, intelligent, smart, quantified via pervasive distributed sensor networks and IOT infrastructures); human augmentation (intellectual, intelligence amplification, prosthesis and exoskeletons); human cognition augmentation for corporate work; corporate work teams of diverse flavors (all-human, all-human with augmented humans, human-machine partnerships and collaborations, human-robot-drone-bot-agent partnerships). Also learn about: extreme, disruptive, data-driven automation, as well as extreme (Big) aspects of the new socio-technical world: scale, volume, velocity, variety and complexity.

The intelligent and smart applications (bots, agents) to be developed and embedded in the Platform Architecture be will used for several goals aligned with several Future of Work (FOW) strategic objectives: instruction, training, gaining experience and expertise in executive management and supervision of future work teams that in various configurations, such as all-human teams with augmented and cognitively enhanced humans, humans with (cognitive, emotional, social, motivational) intelligence amplification, all-robot, all-drone, all-machine, all-bot teams, mixed human-machine teams, hybrids, chimeras and equivalents of colony organisms; continuous, iterative, agile learning; re-training, skill acquisition; gaining understanding and insights into human performance in a sea-of-machines.

In related work, the National Science Foundation (NSF, US) has a funding program in place for research on the implications and applications of Intelligent Cognitive Assistants (ICA) in Human Augmentation for Future of Work, (Oakley, J., (2018).

**Materials and Methods**

The following principles and digital technologies based on them have been integrated to guide how to build the Architecture of the CxO Automation Platforms.

a) Adopt and emphasize Computational thinking and Computational Models of Human Performance Augmentation and Externalized Enhancement. See Table (1) and Fig. (8). The Computational models are to be supported in the Platform as Integrated Technology, and Platform's components, subsystems, modules, serving as agents, bots, software, virtual and digital robots, companions and assistants are themselves, constitutionally cognitive, knowledge-based, smart and intelligent. Computational thinking necessarily results in embodiment in information and software systems of the research ideas, results and deliverables.

b) Focus on externalized (outside-the-body, proxy, delegate, avatar) digital augmentation, enhancement and auxiliary support, (socio-psychology not neuro-psychology).
c) Learning as a form of Innovation, Self-innovation Change. Hence, emphasize several important and effective learning driving aspects such as: active engagement, startup, entrepreneurial thinking, problem-solving, creativity, ingenuity, motivated, goal-oriented and purposive learning and cognitive activities and behaviors.
d) Learning and cognition models that emphasizes Computational thinking (computational learning: computational models of thinking and learning). This also incorporates Design thinking, System thinking, tinkering and Constructionalist learning (S. Papert, J. Piaget).
e) Learning of Content as Content representation as ADT (abstract data types) of knowledge, information, data, intelligence and metadata (database and data) structures; structure traversals, navigation, tours, explorations; and structure manipulations, such as CRUD (create, read, update and delete) operations, further refined into PMSCIO (processing, memory-storage, switching-communication, control-coordination and input-output) operations.
f) Active, autonomous, spontaneous, anticipatory, predictive, speculative computations on intelligent databases (knowledge, information, data, intelligence, metadata DB). That is. Use a system initiative to supplement and support user-driven, on demand, pull interaction with digital companions, agents, assistants and bots (CAB).

Principles for gaining understanding and insights into human cognition and human cognition augmentation:

| Human Performance Augmentation: | Mental, Intellect |
|---|---|
| Cognition augmentation | Thinking, Reasoning, Prediction, Imagination |
| Learning augmentation | Change, accommodation-assimilation (J. Piaget) |
| Motivation augmentation | Influence, persuasion, (R. Cildiani) |
| Emotion augmentation | Disposition to engagement, Socialization, Social intelligence |
| Memory augmentation | Record, Storage, History, Experience, Remembrance, Recall |
| Volition augmentation | Will, Voluntary, Coercion, Compulsion, Addiction |
| Behavior augmentation | |
| Persuasion augmentation | Change (attitude, behavior) (BJ. Fogg, N. Eyal) |
| Attention Control System augmentation | Consciousness, Mind, Self |
| | |
| Cognition augmentation: | |
|     PMSCIO augmentation | (see Table (2)) |
| "Big" (Data) augmentation | |
|     Volume, Scale augmentation | |
|     Variety, Complexity augment. | |
|     Velocity, Speed augmentation | |
|     Veracity, Correctness augm. | EDC/FT |
| "ILities" management augmentation: | |
|     Optimization augmentation | |
|     Meta, Reflection augmentation | |
|     Self-* mgt, Autonomics augm. | |
| Effort SOCAR augmentation: | (N. Lazzaro, gamification: Bartle, J. McGonigal, YK. Chou) |
| Easy fun SOCAR augmentation | SOCAR: separation of concerns, challenges, aspects, roles |
| Hard fun SOCAR augmentation | |
| Serious fun SOCAR augment. | |
| Social fun SOCAR augmentation | |
| | |

Table (1): Aspects of Human Performance Augmentation, Enhance, (Digital) Prosthesis, Exoskeleton, Assistive Technology

| | |
|---|---|
| PMSCIO augmentation | P: processing, M: memory, S: switching-comm, C: control, IO: input/output |
| P augmentation: Thinking augmentation, reasoning (M. Levine): (see also Table (3)) | |
|     Problem-solving thinking augmentation: Heuristics (G. Polya, I. Lakatos) | |
|     Critical thinking augmentation: Analysis \| G.O.D → Selection, Choice \| Scoring, Grading | |
|     Rule-guided thinking augmentation: Procedural, Algorithmic thinking, algorithmic intelligence | |
|     Creative thinking augmentation: design thinking, system thinking, terraforming thinking, agile tinkering, constructionist thinking (J. Piaget, S. Papert, G. Kron) | |
|     Optimization thinking augmentation: satisficing (HA. Simon), Hamiltonian contributions | |
|     Prediction thinking augmentation: speculation, G.O.D, Evolutionary programming | |
| M augmentation: Storage, Representational structures, Preservation, Curation | |
|     Manipulations: Recall, Recollection, Remembering, Reminding, Retrieval, Access, Use | |
|     Manipulations: Structure traversals, navigation, transformation, computation, CRUD | |
|     Representational structures: symbolic, iconic, indexical (semiotics: CS. Peirce) | |
|     Syntactic structures, grammars, language structures, semantics translation structures | |
|     Graphs, Hyper-graphs, networks, polyhedral, graphics, diagram structures | |
|     Spatial, Temporal, Sequential, Mereological, Nested, Hierarchical, Multi-scale structures | |
| SC augmentation: C*: control, coordination, coherence, choreography, cybernetics | |
|     Social intelligence augmentation | |
|     Attention controls augmentation | |
|     Optimization coordination, reduce-assembly augmentation | |
|     Goal-oriented, Purposive, teleological augmentation, automatism, automatics, autonomics | |
|     Stored-program behavior, computing augmentation | |
| IO augmentation: ( see also: NSF FW-HTF Theme 2: Augmentation of Physical Capabilities) | |
|     Visualization augmentation, Data Visualization, STEM visualization | |
|     Virtual, Digital Tangible manipulatives augmentation (H. Ishii) | |
|     AR (augmented reality) augmentation; SDG (synthetic data generation) augmentation | |

Table (2): PMSCIO Computational Model of Thinking-Cognition-Performance Augmentation

| |
|---|
| Dialectical reasoning |
|     Contest, Push-Pull, Ying-Yang struggle, Proofs-and-Refutations (I. Lakatos), Immune sys models |
| Learning-as-Innovation |
|     Change, Persuasion (BJ Fogg, N. Eyal) |
|     Agile, Lean, PDCA reasoning and thinking (Deming, Toyota) |
|     Accommodation-Assimilation (J. Piaget) |
|     Prediction Exploration; Optimization exploration, Colonization |
| Prediction reasoning |
|     Forecasting, Extrapolation, Anticipation (J. Hawkins) |
|     G.O.D (generation of diversity), SDG (synthetic data generation) |
|     Envisioning, Imagining, Imagination, Simulation, Dreaming |
| Exploration reasoning |
|     Discovery, Tour, Navigation, Trail following, Search, Seeking, Wandering-Wondering |
|     Journey, Adventure, Hero's Journey (J. Campbell), Travel, Visit, Traversal |
| Coordination-Control reasoning |
|     Cybernetics, Governance, Regulation, Supervision, Leadership, Rule, Coercion, Compulsion |
|     Homeostasis, Stability |
|     Optimization, Satisficing (HA. Simon), Persuasion (Behavior-Attitude Change), Motivation, Influence, Seduction, Conduction |

Table (3): Some Specialized Forms of Thinking and Reasoning that can be (externally) Augmented and Enhanced

Ultimately, the Architecture integrates several digital technology into digital assistive technology (digital soft prosthesis and exoskeleton), delivered as: (1) intelligent <u>learning companions</u>; (2) work role <u>cognitive companions</u> (assistants, avatars); (3) and <u>server platforms</u>.

The technologies planned for integration include: learning-as-innovation (computational, design, constructionism, systems, agile, lean, and PDCA thinking); assisted deep human learning, (motivation & persuasion engineering, game, gamification systems); self-* automonomics, computational heuristics (G. Polya, I. Lakatos, HA. Simon); cognitive architectures; data-driven computational systems, (soft computing, 5 tribes of algorithms (P. Domingos), Big Data); synthetic data generation and data augmentation, computational manipulations (algebraic, semantic) of attributed structures: (data, information, knowledge, metadata-intelligence) x (spatial, temporal, mereological-multiscale), ER model (hence, semantic networks, semantic web, associative structures, data dependency networks, (multi-dimensional) spreadsheets); visualization and digital visual-tangible-haptic manipulations, augmented reality (AR) based knowledge-information-data-intelligence-metadata manipulations).

Some of the concrete topics and themes to be initially supported: working in corporate teams of human-machine (robot, drone, bot, agent) partnerships; working with all-machine teams; corporate workforce impact of new digital and soft technologies (AI, algorithmic intelligence, IT and ICT).

**Results and Discussion**

The overall architecture of the Platform, as currently conceived, conceptualized and engineered is shown in Fig. (1). The Platform's architecture design, inevitably, will be modified, adapted and re-engineered, based on the results obtained as the research proceeds. What is highlighted is the Platform in its context of a real, actual, authentic corporation, enterprise or institution, subject to the forces involved in the interaction of digital technology and future of work. The real enterprise itself, (geographically and potentially globally, spatially and geographically distributed), exists in the context (ambience or surrounding) of the external physical environmental, ecological and societal-social-political-economic environment, world or space.

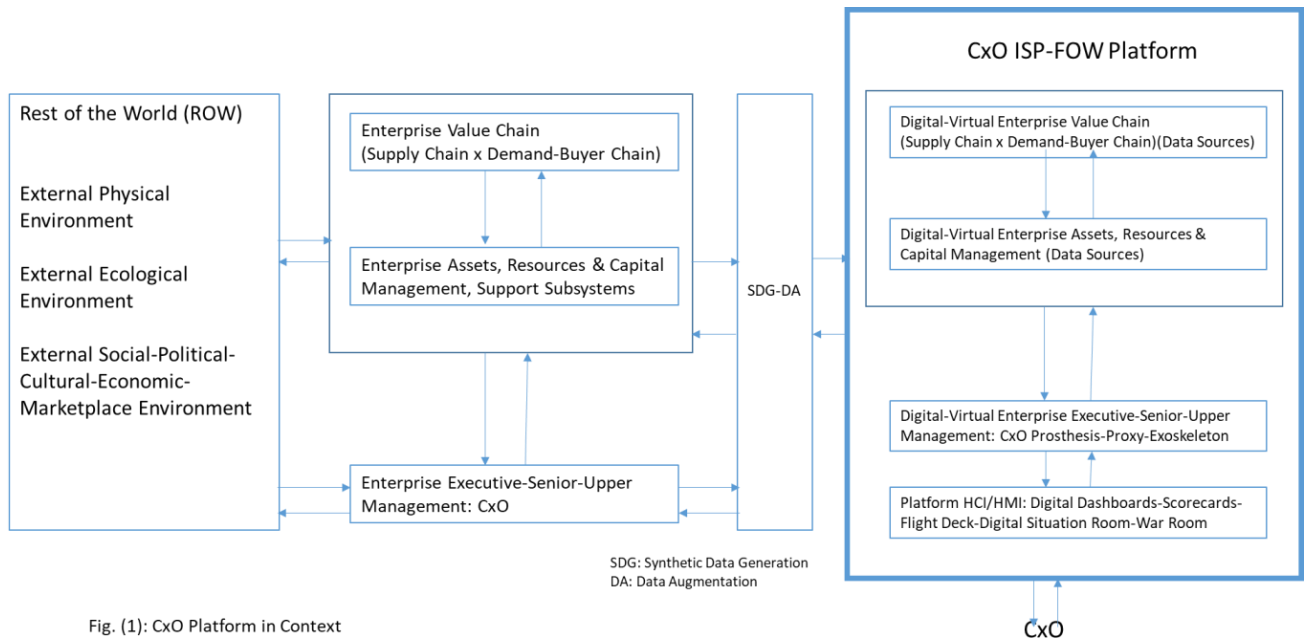

Fig. (1): CxO Platform in Context

The Platform itself is a digital world isolated, segregated and separated from the physical enterprise with which it's associated or coupled; hence a virtual, digital micro-world. Data generated by the physical enterprise (in most instances in the form of massive scale Big Data) will be injected into the Platform, after pre-processing.

Since the Platform embodiments and deployable structures will be digital micro-worlds, a key and essential aspect of the proposed research is the use of Synthetic Data Generation (SDG) and Data Augmentation digital technologies to provide realistic Big Data sources and data streams that drive the intelligent agents organized and incorporated into the Platform. Essentially, real, authentic data from the associated enterprise will be manipulated using digital technology to obtain (virtual, synthetic, fake, counterfeit) data with some specified measure of realism, authenticity and credibility that can serve as that data sources for the Platform's sensors, detectors, meters and instruments that perform is data collection, gathering, acquisition, capture, and harvesting. Synthetic data can also be generated from relevant sources in the external context of the enterprise, (matching marketplaces, dual sided and multi-sided platforms, smart city IOT, transportation IOT, utilities IOT, public health IOT, and healthcare IOT).

Some What-if scenarios the Platform is designed to facilitate include:

    Corporate mergers and acquisitions

- Corporate re-organizations, re-engineering, evolution, upgrades, transitions and migrations

- Employee and workforce (skillset) digital augmentation

- Managing, supervising all-robot, bot or digital agent teams for work projects, missions, tasks.

- Using autonomous robots for middle management, especially, digital agents managing human-robot-agent-bot teams

- Integrating disruptive technology into enterprise and corporate strategy, plans, tactics and pragmatics.

- Employee and workforce adaptation to (digital) technology, including scenarios of replace, substitution or usurpation of roles vs. augmentation, cooperation and collaboration, (where to work, where to conduct work, when to work, how to work, how often to work, which work, what work, how long to work, who to work for, self or other; why to work).

- "iLities" Management and Analytics: impacts on productivity, throughput, efficiency, efficacy, quality, safety, security, convenience

- Employee adaptation and learning: hard STEM skills; soft social intelligence, emotional intelligence and interpersonal skills.

Queries, questions, analysis and what-if scenarios to gain insight about impacts on technology on future of work are ultimately concern business and corporate roles and role inter-relationships and associations Therefore, computational models are required and needed to represent corporate roles. The Platform Architecture uses attributed structures to computationally model corporate roles and role complexes. Attributed structures are extensions of attributed parse trees and syntax trees of attribute grammars (2-level grammars) (D. E. Knuth). Topologically, the role models can be spatial structures (graphs, hyper-graphs, networks, tessellations, tiling, honeycombs, polyhedral, polytopes of combinatorial topology), temporal diachronic structures (sequences, streams, time series, lifestreams), and mereological structures (multi-scale, nested hierarchies, heterarchies and holarchies). Algebraically, these structures can be symbolized as elements of scales of sets (N. Bourbaki), which equivalent to Grassmann multiple quantities that form the hierarchy or staircase of scalars, vectors, arrays, tensors and holors.

As a further refinement, each role, as an individual node in the computational structures noted above can be represented using the (entity-relationship) ER model. Each role is entity or relationship object, with associated attributes (properties, features, properties, qualities), and attribute values. With such computational models, termed the ADAAM technology (augmented data and augmented metadata), corporate roles can be analyzed, and changes, impacts or effects, determined using of comparisons of human-based roles and machine, automation and digital agent technology based roles.

The internal logical architecture of the Platform can be design to mirror and reflect the internal details of the generic LSE or MNC enterprise. Suitable model for representing such is the Value Chain Model, which consists of a coupling of a Supply Chain Model and a Demand Chain Model [Michael Porter]. A further refinement of the Value Chain Model is shown in Fig. (2), again based on [Michael Porter], but extended to be a Dual Process Model architecture, wherein one sub-system is quick and automatic, while the other sub-system is slow, reflective and autonomic (involved with meta, self-* management concerns, aspects and roles).

| PPP | IBL | PMP | OBL | WDD | SMU | OUS |

(a): Enterprise Value Chain

| ACS | ACS | ACS | ACS | ACS | ACS | ACS |
| PPP | IBL | PMP | OBL | WDD | SMU | OUS |

(b): Augmented Enterprise Value Chain

Value Chain stages, phases:
    PPP: Procurement, Purchasing, Provisioning
    IBL: InBound Logistics
    PMP: Production, Manufacturing, Process
    OBL: Outbound Logistics
    WDD: Warehousing, Depository, Distribution
    SMU: Sales, Marketing, Use
    OUS: Operations, Use, Service

ACS: Auxiliary, Context, Support:
 Auxiliary Support: Assets, Resources, Capital
    Facilities, Engineering, Technology, Equipment, Factories, Labs
    HR, Employees, Workforce, Labor
    Finance, Accounting, Capital, Credits, Stock Valuations
Context:
    Strategic partners, Allies
    Suppliers, Service Providers, Contractors
    Customers, Clients, Consumers, Users
    Markets, Marketplaces, Matching Marketplaces, Multi-sided Platforms (MSP)
    Competitors, Rivals, Substitutes
    Community, Public, Society, Institutions
    Government, Regulators, Legislators

Fig. (2): Enterprise Value Chain

The enterprise structure at this scale can be re-interpreted as a pipelined computer architecture, with aspects which include multi-scale, super-scalar and multi-pipeline or poly-pipeline features.

The implementation embodiment can use a service oriented architecture (SOA) framework, employing one or more enterprise system buses (ESB) to organize multi-agent system (MAS) servers, services, digital ants: workers, bots, agents, chatbots, digital assistants, which are integrated, inter-coordinated, choreographed and orchestrated into star schema and snowflake schema (tree, forest) architectural patterns, see Fig. (3).

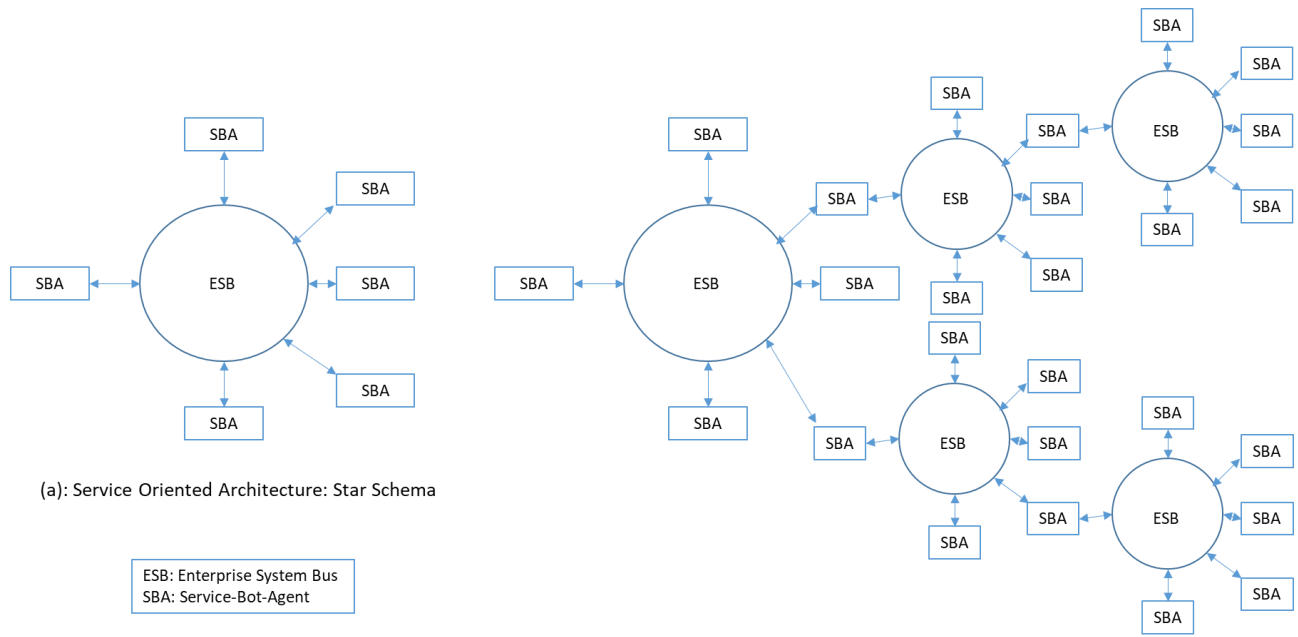

Fig. (3): CxO Tech Intelligence Platform SOA

(a): Service Oriented Architecture: Star Schema

ESB: Enterprise System Bus
SBA: Service-Bot-Agent

(b): Service Oriented Architecture: Snowflake, Tree Schema

The intended and anticipated use of the CxO Automation FOW-ISP Platform can be appreciated by considering the platform as a digital technology, and IT (hardware and software testbed) equivalent of flight simulators, wind tunnels and micro-worlds. Existing, putative, potential, aspiring CxO's and students (particularly from female, minority and disadvantaged populations) can use the Platform for training, education, learning and teaching (TELT) activities. The Platform will be designed the following activities:

(a) Exploration: Tasks will include: discovery of impacts, effects, what-if scenarios and cases, simulations, sandbox play and serious games (simulations, games, modeling + play, animations and visualizations (SGM+PAV)). Other learning opportunities will include investigating and querying about CxO employee augmentation and intelligence amplification (cognitive, skills, emotional, affective, motivational, behavioral), as well as the role of persuasion technology for attitude and behavior change. Another set of tasks is using the Platform to investigate digital agents, bots and digital ants, augmented reality (AR) applications and corporate management digital avatars, that can serve as CxO executive digital prosthesis, exoskeleton, intelligence amplifier and intelligence augmentation.

(b) Practice on coping with emergencies, disasters (EDC) and crisis management and fault-tolerance (EDC/FT), in the era and context of digital technologies and future of work. Tasks and scenarios will include mitigation, short term (tactical) responses, and long (strategic) recovery, as well as EDC prevention and avoidance. This aspect will also include rehearsals and dramatizations, virtual or digital boot camp for CxO, training and instruction for "decision making, planning, and collaboration."

(c) Learning about long term evolution and trends of digital technology and future of work. Platform users will be able use the Platform as a safe digital micro-world to create scenarios and cases that result in corporate and organizational failures, particularly those due to "big data" features of complexity, variety, volume, scale and velocity or speed. The Platform will provide opportunities to CxO's to gain familiarity, experience, expertise and deep human learning about the clash of digital technologies and the future of work.

It is anticipated that the deployable versions of the Platform shall be integrated, coordinated, orchestrated and choreographed (colony organisms, super-organisms, super-craft) of ensembles and collectives of modular intelligent, smart and digital agents, bots, automata and machines (digital ants). The Platform's dynamic assemblage structures will be architected and designed to be reconfigurable, re-programmable, re-usable, multi-scale and compositional.

The human-machine (human-computer, human-digital agent) interfaces of the Platform are designed to be CxO-targeted, domain-specific and application-specific or task-specific digital dashboards, scorecards, and analytic applications for business intelligence.

Some What-if scenarios the Platform will be designed to facilitate include:

    Corporate mergers and acquisitions

    Corporate re-organizations, re-engineering, evolution, upgrades, transitions and migrations

    Employee and workforce (skillset) digital augmentation

    Managing, supervising all-robot, bot or digital agent teams for work projects, missions, tasks.

    Using autonomous robots for middle management, especially, digital agents managing human-robot-agent-bot teams

    Integrating disruptive technology into enterprise and corporate strategy, plans, tactics and pragmatics.

    Employee and workforce adaptation to (digital) technology, including scenarios of replace, substitution or usurpation of roles vs. augmentation, cooperation and collaboration, (where to work, where to conduct work, when to work, how to work, how often to work, which work, what work, how long to work, who to work for, self or other; why to work).

    "iLities" Management and Analytics: impacts on productivity, throughput, efficiency, efficacy, quality, safety, security, convenience

    Employee adaptation and learning: hard STEM skills; soft social intelligence, emotional intelligence and interpersonal skills.

Since the Platform embodiments and deployable structures will be digital micro-worlds, a key and essential aspect of the proposed research is the use of Synthetic Data Generation (SDG) and Data Augmentation digital technologies to provide realistic Big Data sources and data streams that drive the intelligent agents organized and incorporated into the Platform, (Fig. 4). Essentially, real, authentic data from the associated enterprise will be manipulated using digital technology to obtain (virtual, synthetic, fake, counterfeit) data with some specified measure of realism, authenticity and credibility that can serve as that data sources for the Platform's sensors, detectors, meters and instruments that perform is data collection, gathering, acquisition, capture, and harvesting. Synthetic data can also be generated from relevant sources in the external context of the enterprise, (matching marketplaces, dual sided and multi-sided platforms, smart city IOT, transportation IOT, utilities IOT, public health IOT, and healthcare IOT).

The Platform itself is a digital world isolated, segregated and separated from the physical enterprise with which it's associated or coupled; hence a virtual, digital micro-world. Data generated by the physical enterprise (in most instances in the form of massive scale Big Data) will be injected into the Platform, after pre-processing. This is an architectural realization of the Digital Twin concept, (Datta 2016, Marr 2017, Grieves 2014).

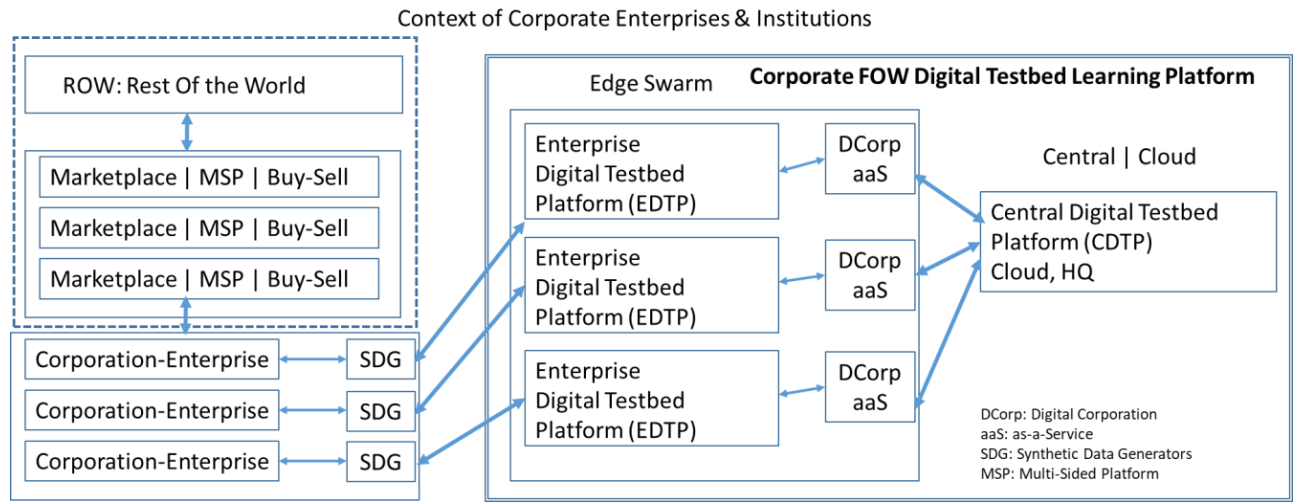

Fig. (4): CxO Automation Platform as Operational Deployable Architecture

Since the Platform embodiments and deployable structures will be digital micro-worlds, a key and essential aspect of the proposed research is the use of Synthetic Data Generation (SDG) and Data Augmentation digital technologies to provide realistic Big Data sources and data streams that drive the intelligent agents organized and incorporated into the Platform. Essentially, real, authentic data from the associated enterprise will be manipulated using digital technology to obtain (virtual, synthetic, fake, counterfeit) data with some specified measure of realism, authenticity and credibility that can serve as that data sources for the Platform's sensors, detectors, meters and instruments that perform is data collection, gathering, acquisition, capture, and harvesting. Synthetic data can also be generated from relevant sources in the external context of the enterprise, (matching marketplaces, dual sided and multi-sided platforms, smart city IOT, transportation IOT, utilities IOT, public health IOT, and healthcare IOT).

Digital Twin Architecture

The enterprise structure at this scale can be re-interpreted as a pipelined computer architecture, with aspects which include multi-scale, super-scalar and multi-pipeline or poly-pipeline features.

The implementation embodiment can use a service oriented architecture (SOA) framework, employing one or more enterprise system buses (ESB) to organize multi-agent system (MAS) servers, services, digital ants: workers, bots, agents, chatbots, digital assistants, which are integrated, inter-coordinated, choreographed and orchestrated into star schema and snowflake schema (tree, forest) architectural patterns, see Fig. (3).

The underlying Project is also studying augmentation of human cognition to support corporate employee learning, (Fig 5), knowledge acquisition and expertise development for re-skilling preparedness for future of work in the next decade. The initial scope of the content to be learned shall be limited to a) new digital technologies (characterized by being advanced, intelligent, smart, cognitive, AI-driven, data-driven, using algorithmic intelligence, and computational intelligence, disruptive, pervasive and ubiquitous); b) human cognition augmentation as assistive technology, and human physical abilities augmentation using smart, intelligent digital prostheses and exoskeletons; c) use of human cognition augmentation in corporate work roles; d) future human roles in corporate future of work (FOW) that utilizes all-machine teams, or hybrid cohorts based on human-machine partnerships, collaborations and cooperatives.

The FOW learning application of the Platform has the following focus:

1. Assessing and evaluating the learning environment, platform performance, effectiveness and efficiency.
2. Student learning performance.
3. Student leverage of learning: retention, transfer, creativity.
4. Effectiveness of using the Platform as opportunity for aspirants and students who are women, minorities and members of disadvantaged groups, to learn, train and become experienced with working and managing the new workforce and workspaces.

Consequently, the CxO Automation FOW Platform can be utilized to achieve the following aims:

1. Advance Human Cognitive Capabilities, via lifelong learning on an intelligent platform of learning companions.
2. Augment the learning aspect of human cognition, with learning companions of the platform as externally-resident digital intelligence amplifiers, prosthesis, exoskeletons.
3. Augment human performance, workplace skillsets, deep learning about new, intelligent, digital technologies as disruptive influences.

The human-machine (human-computer, human-digital agent) interfaces of the Platform is designed to be CxO-targeted, domain-specific and application-specific or task-specific digital dashboards, scorecards, and analytic applications for business intelligence.

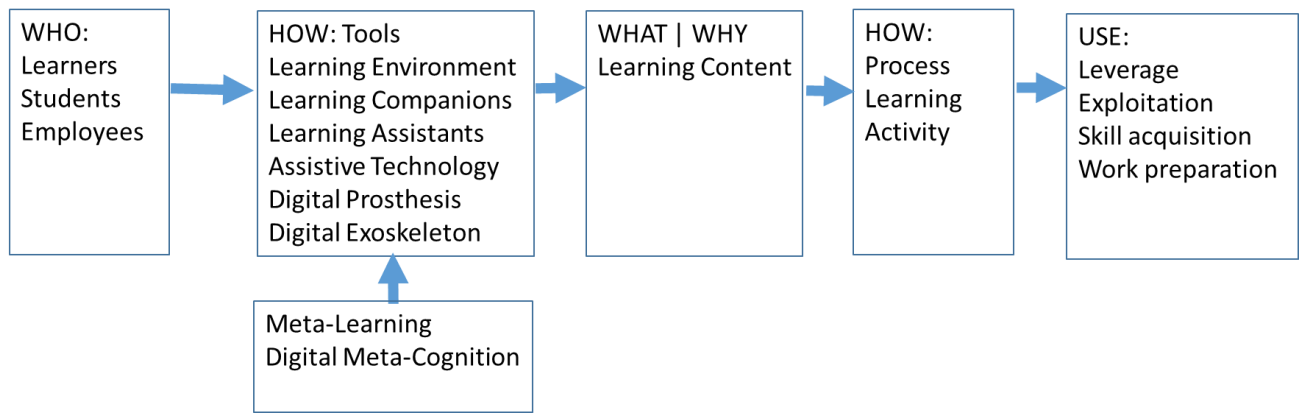

Fig. (5). Learning Environment for Human Cognition Augmentation

The platform can be extended for dynamic, continuous, lifelong learning of human cognition augmentation in the context of future of work.

The following specific types of fundamental scientific research based on the Platform:

* Assessing and evaluating the learning environment, platform performance, effectiveness and efficiency.

* Student learning performance in knowledge (learning content) acquisition: velocity, variety, complexity, volume and scale, correctness and integrity.

* Student leverage of learning: retention, transfer, creativity.

* Effectiveness of using the Platform as opportunity for aspirants and students who are women, minorities and members of disadvantaged groups, to learn, train and become experienced with working and managing the new workforce and workspaces.

* Collaborative research on Project multi-disciplinary impacts: corporate economic, social-psychological, STEM.

* Potential impact of Project's platform on nation's economic performance.

Lifelong learning can be modeled as continuous total quality improvement. Thus, one can support such learning by emphasizing strategies and techniques of PDCA (Plan-Do-Check-Act) (Deming), Agile tactics, Lean thinking, Lifestream-as-ADT manipulation (Gelernter), continuous heuristics (G. Polya, I. Lakatos).

**Summary and Conclusions**

Automation of CxO functions needs to be addressed as part of the disruptive impacts of advances in digital technologies on Future of Work. Several design and engineering principles, including modularity, multi-scale structure, digital twin, ADT manipulations based computational thinking, can be used to create the complex architecture required for a platform that can be used to implement and embody CxO automation.